\documentclass{sigchi}

\usepackage{balance}  % to better equalize the last page
\usepackage{graphics} % for EPS, load graphicx instead 
\usepackage[T1]{fontenc}
\usepackage{algpseudocode}
\usepackage{txfonts}
\usepackage{mathptmx}
\usepackage[pdftex]{hyperref}
\usepackage{color}
\usepackage{booktabs}
\usepackage{textcomp}
\usepackage{microtype} % Improved Tracking and Kerning
\usepackage{ccicons}  % Cite your images correctly!
\usepackage{todonotes}

\def\plaintitle{Instagram Post Data Analysis}

\def\emptyauthor{}
\def\plainkeywords{data visualization, instagram, filter, hashtag}

\makeatletter
\def\url@leostyle{%
  \@ifundefined{selectfont}{
    \def\UrlFont{\sf}
  }{
    \def\UrlFont{\small\bf\ttfamily}
  }}
\makeatother
\def\pprw{8.5in}
\def\pprh{11in}

\definecolor{linkColor}{RGB}{6,125,233}
\hypersetup{%
  pdftitle={\plaintitle},
  pdfauthor={\emptyauthor},
  pdfkeywords={\plainkeywords},
  bookmarksnumbered,
  pdfstartview={FitH},
  colorlinks,
  citecolor=black,
  filecolor=black,
  linkcolor=black,
  urlcolor=linkColor,
  breaklinks=true,
}

\begin{document}

\title{\plaintitle}

\numberofauthors{1}
\author{%
  \alignauthor{Steve Chang\\
    \affaddr{University of California, Los Angeles}\\
    \email{USA}}\\
}

\maketitle

\begin{abstract}
Because of the spread of the Internet, social platforms become big data pools. From there we can learn about the trends, culture and hot topics.  This project focuses on analyzing the data from Instagram. It shows the relationship of Instagram filter data with location and number of likes to give users filter suggestion on achieving more likes based on their location. It also analyzes the popular hashtags in different locations to show visual culture differences between different cities.
\end{abstract}

\category{H.5.m.}{Information Interfaces and Presentation
  (e.g. HCI)}{Miscellaneous}{}{}

\keywords{\plainkeywords}

\section{Introduction}

As "a picture is worth a thousand words", more and more people are sharing their daily life, personal interests, news and events using images on social platform. Instagram is such a platform which is a popular mobile photo sharing application. It launched in October 2010, and rapidly gained popularity, with over 100 million active users as of April 2012 and over 300 million as of December 2014. Because of these huge amount of information in the dataset of Instagram, and we are interested in analyzing and visualizing the patterns of them. 

This project is focusing on analyzing Instagram data to learn about the culture differences between different places. This project analyzes how filter usage are distributed in 50 cities, which are the cities with most population in each state of United States. This give us the information about how filter preference and vision culture varies in different states. The project also analyzes the number of hashtags been labeled on the posts for each city. It shows the popular hashtags for each city which reveal the popular event or hot terms in different cities.

While you are using the Instagram, you may also have ever scrolled through the Instagram filter list back and force worrying about which one to use, and how to make more people like it. But since culture background and contents varies a lot from photo to photo, it is hard to make a simple suggestion that let everyone like it. Our project also analyzes the Instagram filter data based on location and like to help you solve this problem.

The goal of this project is to learn about visual culture and content differences to help catch both the artistic trends and event trends for different places. It can also help user to make better filter selection to improve photo quality and reach more likes.   

\begin{figure}[ht]
  \centering
\includegraphics[width=3in]{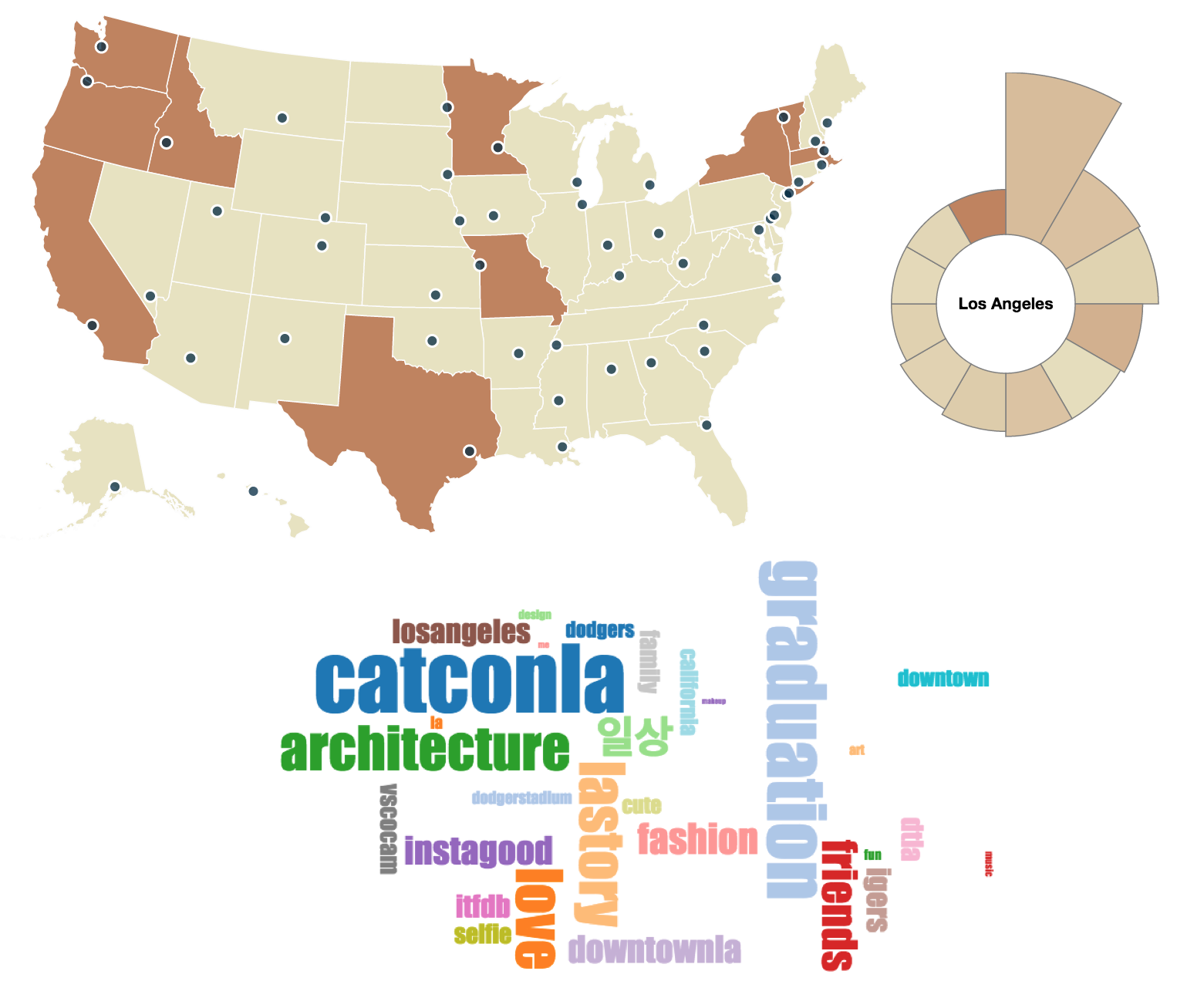}
  \caption{Visualization system design}
  \label{like-filter}
\end{figure}

\section{Related Work}

There are many kinds of visualization, from a monitoring tool to visualize pressure map \cite{huang2014using}, to an interactive tool to visualize Twitter data \cite{wang2015senticompass}. Due to the popularity and the amount of data information, there are couple groups of people are also interested in the culture information reflected in the Instagram photo data. Hochman's Zooming into an Instagram city \cite{hochman13} \cite{hochman12} visualizes and analyzes samples from a data set of about 550,000 Instagram photos from New York City and Tokyo, by applying visualization and Cultural Analytics techniques. They show all the images in the collection. They download data based on latitude and longitude criteria and use specified tools to analyze the data. One interesting result of this paper is that there seem to be reoccurring spatio-temporal visual deviations in a specific time period and a set place. Based on those large sets of Instagram photos, they show how visual social media can be analyzed at multiple spatial and temporal scales. They also present analysis of social and cultural dynamics in specific places and particular times, and introduce new visualization techniques which can show tens of thousands of individual images sorted by their metadata or algorithmically extracted visual features. But as they only focuses on photo color and photo style, but our project cares more anout the label information and filter data. They shows the differences of visual styles in different time different cities, while we are analyzing the exact events and objects that people are captured and marked in different cities.

There are also some existing analytical tools for Instagram. Iconosquare (formerly Statigram) \cite{iconosquare} provides useful statistics about Instagram. It can also respond to comments and monitor hashtags. Instagram-analytics \cite{simplymeasured} contains huge amount of raw instagram data for users. Instastats \cite{instastats} is python scripts to pull data from Instagram API. Those work shows us what data information are available from Instagram and our project uses some of the tools to get our desired data for analysis.

\section{Methods}

We implement a software tool for user to navigate the Instagram data. The implementation details are divided into four parts:

Data preparation: getting data from Instagram and sorted our desired data into useable csv format.

Data analysis: we use table and chart to pre-analyze the data.

Function design: our system filters data based on location, and visualizes filter and tag data.

Visual style design: our system uses map to display location data, uses Aster chart to visualize filter data, and uses word cloud to visualize hashtag data.

\subsection{Data Preparation}

First, we get Instagram data from Instagram API which allows us to specify the scope of the access we are requesting from the user. It provides four types of scopes:

basic - to read data on a user's behalf, e.g. recent media, following lists (granted by default);

comments - to create or delete comments on a user'��s behalf;

relationships - to follow and unfollow accounts on a user's behalf;

likes - to like and unlike media on a user's behalf.

In our project, we are insisting in the basic and likes scopes including filter types, tags, likes, image link and location information. We get the data through API endpoint and use instastats python script to pull data. The example data is shown below:

"data": [\{\\
\text{\quad}        "filter": "Earlybird",\\
\text{\quad}        "tags": ["expobar"],\\
\text{\quad}        "likes": \{\\
\text{\quad \quad}              "count": 35,\\
\text{\quad}        \},\\
\text{\quad}        "link": "http://instagr.am/p/BUS3X/",\\
\text{\quad}        "location": \{\\
\text{\quad \quad}            "latitude": 37.780885099999999,\\
\text{\quad \quad}            "longitude": -122.3948632,\\
\text{\quad}        \}\\
    \}]

We found data collection process is painful. Instagram's API has unpredictable behavior. For example, even if we indicate the number of post we want in query, its API will still give us different number of result in each run. Thus, in order to get enough data, we need to run multiple times and remove duplicate results.

\subsection{Data Analysis}

\begin{figure}[ht]
  \centering
  \includegraphics[width=3in]{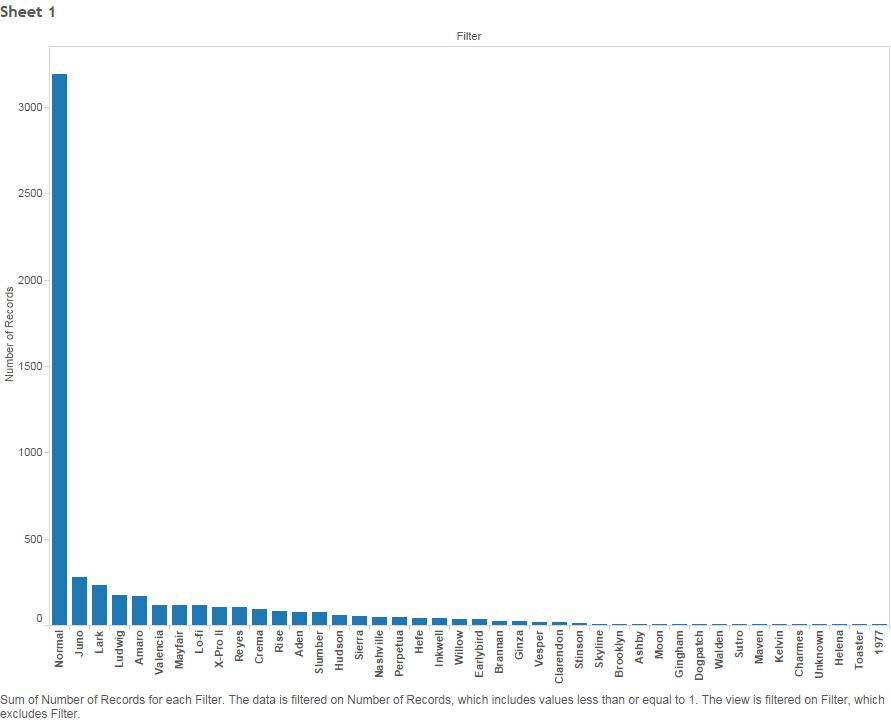}
  \caption{Number of photos in different filters}
  \label{normal}
\end{figure}

\begin{figure}[ht]
  \includegraphics[width=3in]{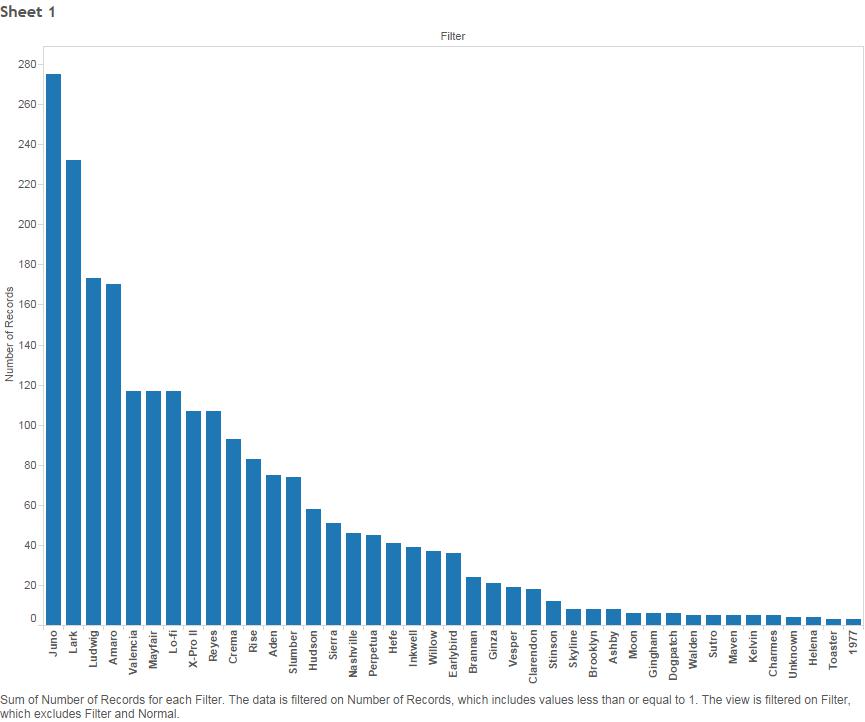}
  \centering
  \caption{Number of photos in different filters. Normal excluded}
  \label{no-normal}
\end{figure}

In order to find interesting pattern from the raw data, we use python script to process the data. We first explore the relationship between filter and number of likes. We ignore the location and get the whole table of filter, number of photos with this filter, maximal likes, average likes and total number of likes.  With the help of Tableau, we can see the following result in figure \ref{like-filter},\ref{normal} and \ref{no-normal}.

\begin{figure}[ht]
  \centering
\includegraphics[width=3in]{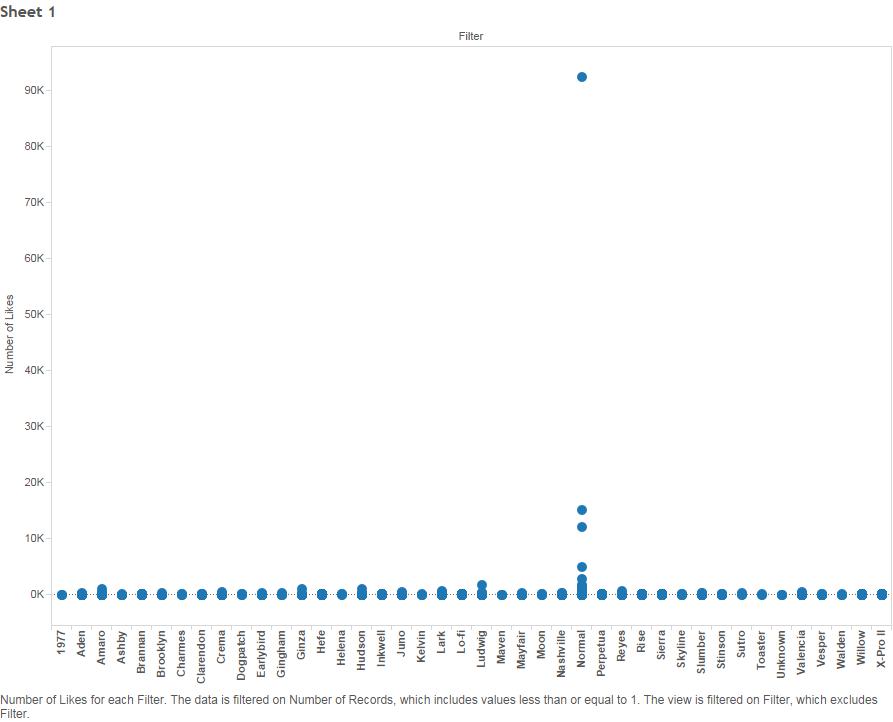}
  \caption{Likes and filters in Seattle}
  \label{like-filter}
\end{figure}


We can see that the normal filter appears dominantly. This is probably because normal filter is the default setting of Instagram. Therefore if people are not very familiar with Instagram, then they will hardly pick up advanced filters. So many filters use normal filter that most popular photos are using normal filters. If we remove the effect of normal filters, we can see that the curve of  number of photos on filters looks like an exponential function, which is an evidence that it may  follow the power law...(line truncated)...
In this way, we may design a recommendation system for filters:
since normal filter is so common,
we can help users to use other advanced filters,
in this way users may obtain sense of accomplishment while achieving lots of likes.

We also study the relationship between the likes and the hashtag. We create the table of word, total like, average like and maximal like for each city. The following chart is the result for Seattle.
\begin{figure}[ht]
  \centering
  \includegraphics[width=3in]{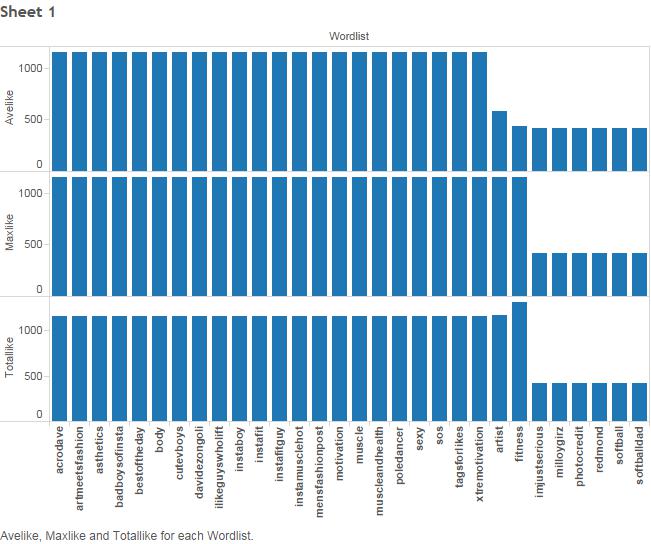}
  \caption{Hot hashtags in Seattle}
\end{figure}
We do not find interesting pattern in the chart. We think the main reason is that we do not have enough data. Most hashtag appear only 1 or 2 times, which is too few to analyze. However, we can still observe that hashtags with popular photos are “meaningful”, that is , we can see some kind of trend from the hot hashtags. This can be used to notice users what is happening now in the city.

\subsection{Function Design}
There are three main functionality, focusing on location, filter and tag data.

\subsubsection{Location}
User can click on dark blue dot, which is the city with largest population in a state. Then, the system shows the filter and tag visualization result based in the specific location. We want to use location as a filter for visualization results.

\subsubsection{Filter}
In the filter visualization result, we want to show the average likes of photos using these filters. We also want to show how many photos using those filters, which indicates the popularity of each filter.

\subsubsection{Hashtags}
In the tags visualization result, we want to show the average likes of photos using these tags.

\subsection{Visual Design}

We design our tool color theme using Instagram color patterns since we are analyzing Instagram data, this can provide users a more related and clear connection between the data information and the tool. At the top of the page it shows a brief introduction of the project and overall functions to gives users a understanding about how to navigate and explore the tool.

\subsubsection{Location - Map}

\begin{figure}[ht]
  \centering
\includegraphics[width=3in]{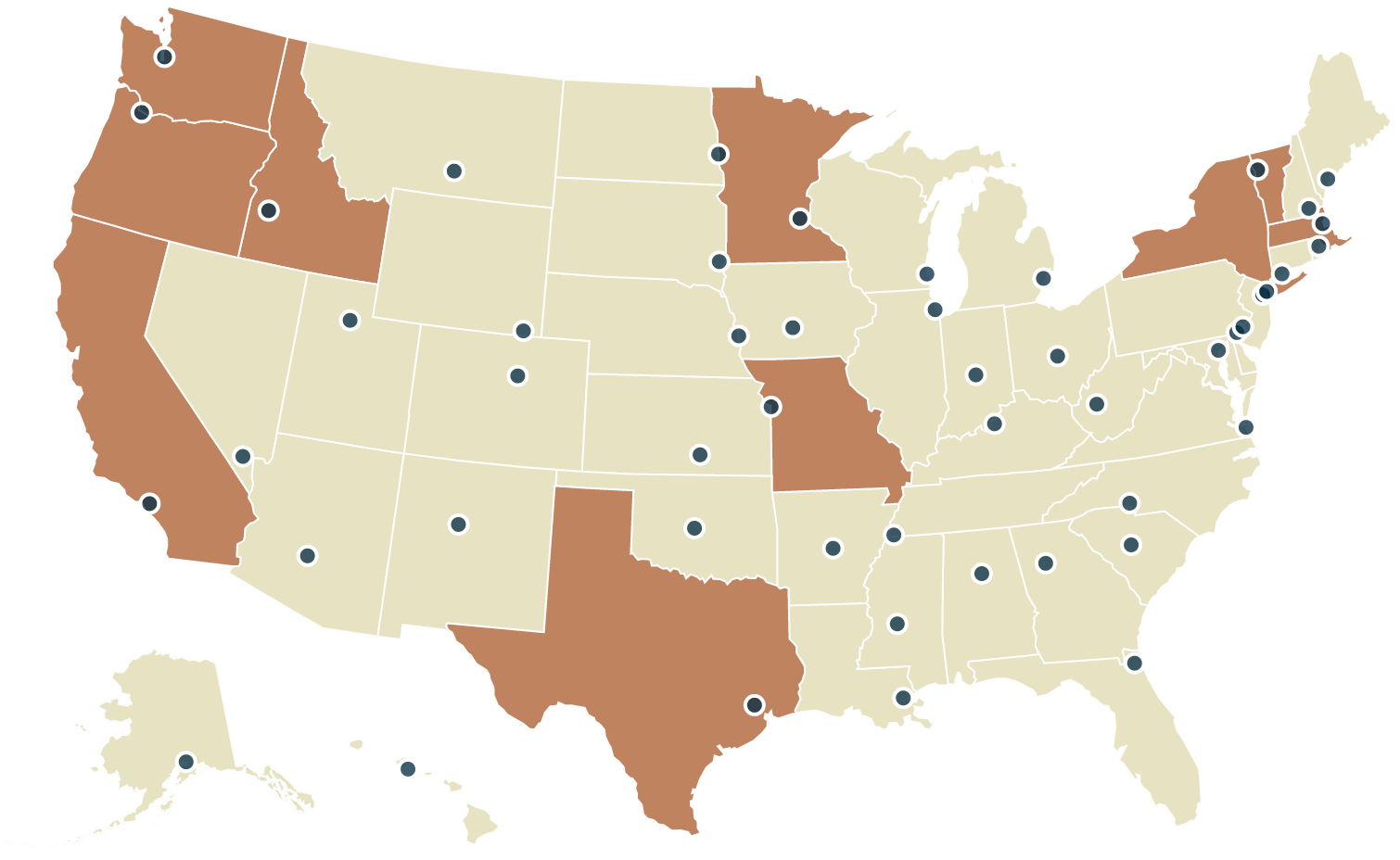}
  \caption{Map design}
  \label{like-filter}
\end{figure}
The project uses the map of United States. The color of the map is also chosen based on the Instagram color, the states with darker brown color are the top ten most Instagram posts states in the given time period, which is the recent past week. Lighter yellow color are the rest states. Since darker color presents more and higher density, so we choose darker color to represent more posts. Using dark blue color as the city dot is to form a contrast with the yellow and match the Instagram color theme.

\subsubsection{Filter - Aster Chart}

\begin{figure}[ht]
  \centering
\includegraphics[width=3in]{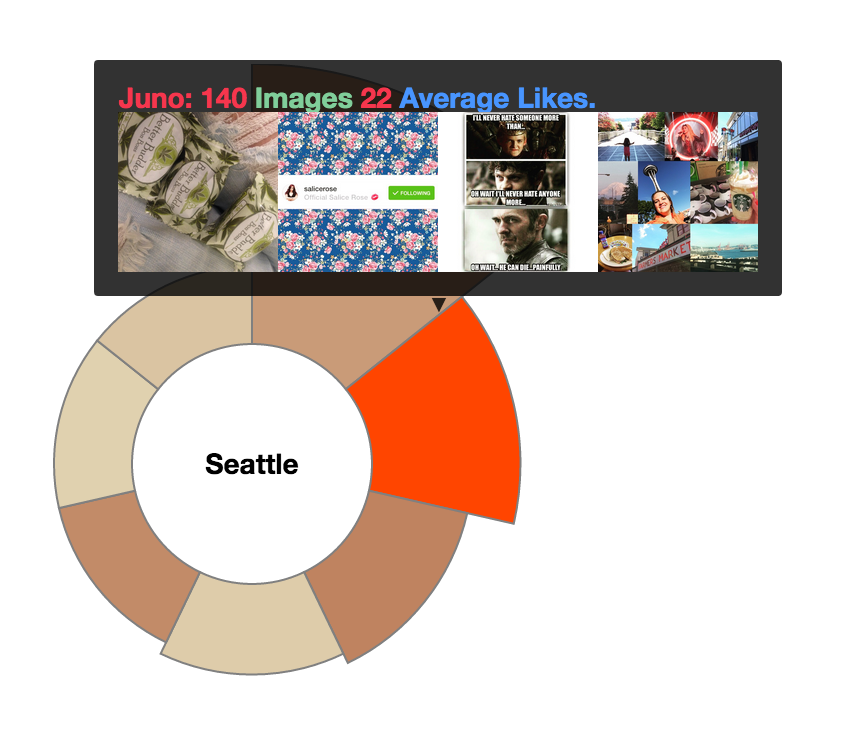}
  \caption{Filter chart design}
  \label{like-filter}
\end{figure}

When click on the city dot, an Aster chart\cite{aster} that shows the relationship between the filters and likes for the selected city shows up. Larger slice means more posts for that filter and smaller slice means less posts for that filter, because the size can be related to the amount information. As for the color, lighter color of the chart slice means less likes for that filter, and darker color means more likes for that filter. Most liked one uses the brown color and the least liked one uses the yellow color where the colors are match the color of the map. The color for the rest filters are the color within this range and equally scaled. When mouse hovering on the slice, the slice turns into red color to use contrast color to indicate one filter type is selected, and a window pop up near by the mouse showing the detail information with the number of posts and likes and sample images for selected filter type. The color of the text information are using couple bright colors selected from Instagram logo which can gives the sense of users that they are viewing Instagram posts and draw some attention on the text because of the brightness. 

\subsubsection{Hashtags - Word Cloud}
\begin{figure}[ht]
  \centering
\includegraphics[width=3in]{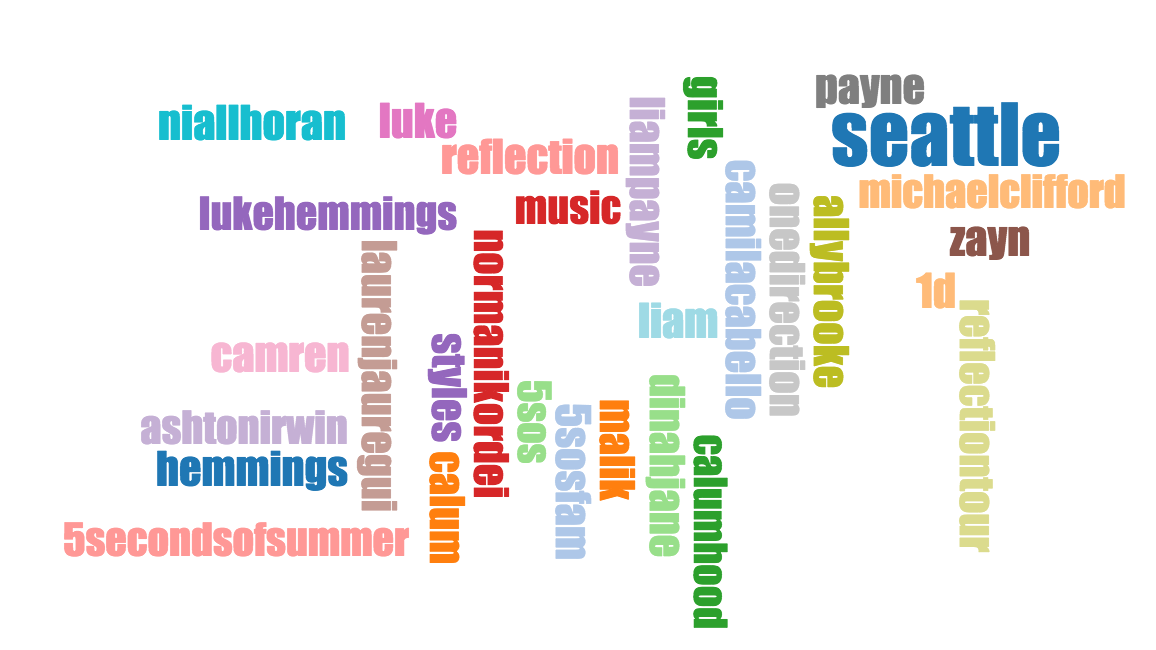}
  \caption{Word cloud design}
  \label{like-filter}
\end{figure}
After clicking the city dot, there is also a words cloud\cite{word} with tags showing underneath the map. Larger size of text means more posts with that tag and smaller size of text means less posts with that tag. The color of the text are all bright colors with large contrast to distinguish words from words and also for user easy to read.

\section{Results}
Our final work can be found in http://xiaoyizhang.me/512 . Basically, the top 10 states with most instagram posts can be seen directly as the darker parts on the map. Users can choose one city to see the hot hashtags people are  using. Also users can see the amount and the popularity of different filters.

For the choice of filters, as we have mentioned, normal filter is most frequently used. In every city, the number of photos in  normal filter is time times of any other filters. However, with respect to average, normal filter does not win everywhere. For example, in  Charleston, the Lark filter achieves more average likes. So it is possible for advanced users to obtain better effect with interesting filters. 

\begin{figure}[ht]
  \centering
\includegraphics[width=3in]{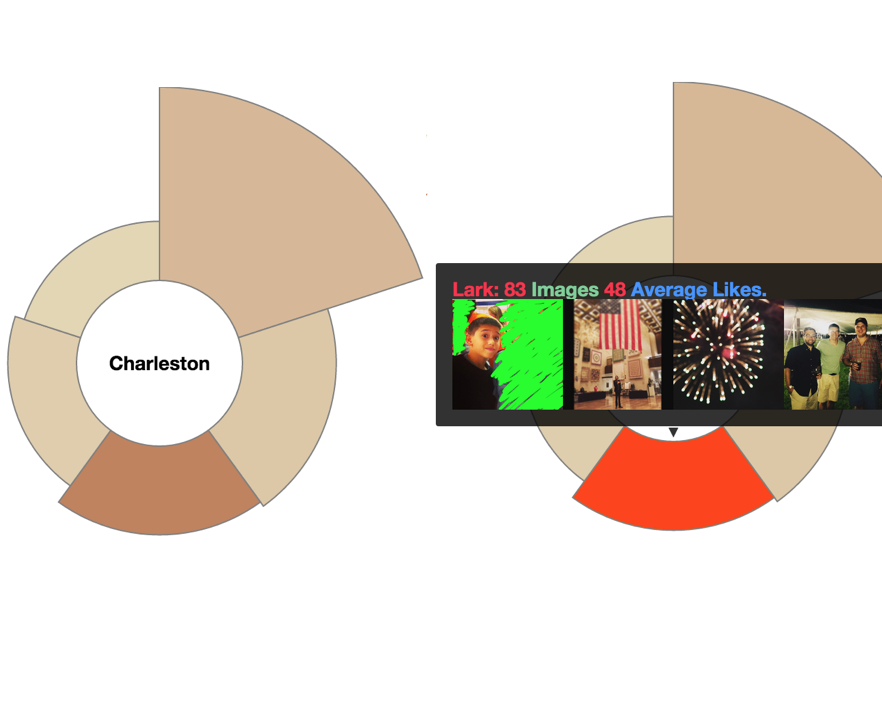}
  \caption{Illustration for Charleston}
  \label{like-filter}
\end{figure}

For the hot words, we can observe culture difference between different cities, Of course, the name of the location is a quite popular hashtag; however, if you check the word cloud of LA, you can find "graduation" and "fashion", which implies LA is a energetic city; on the other hand, for New York, you can see the names of a lot of Japanese animations. One possible explanation is that New York may be holding an anime exposition at that time..  

\begin{figure}[ht]
  \centering
\includegraphics[width=3in]{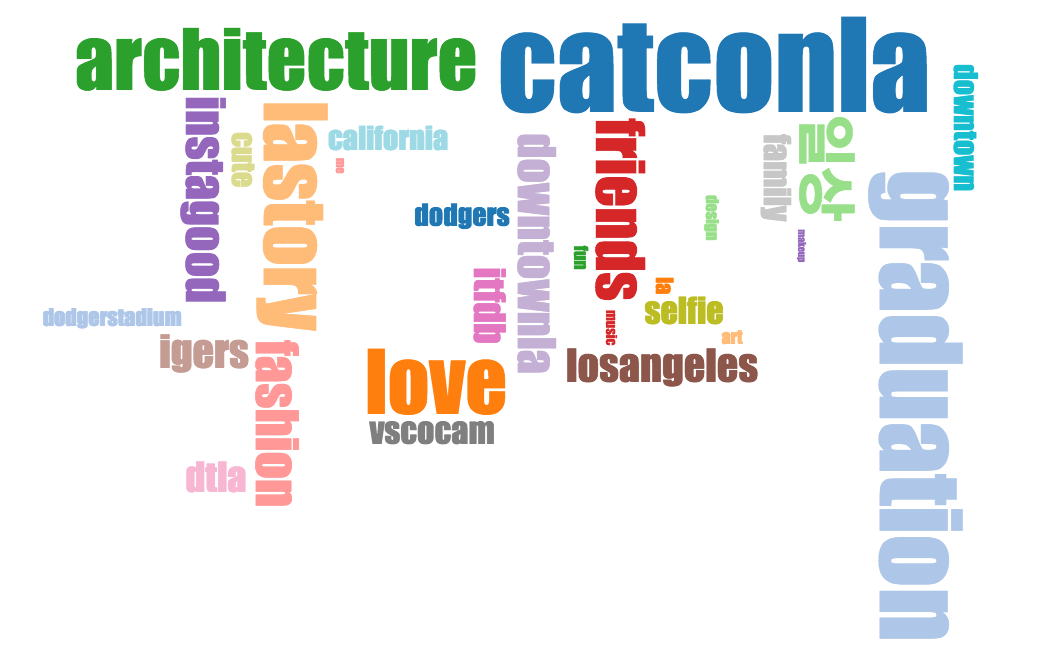}
  \caption{Illustration for LA}
  \label{like-filter}
\end{figure}

\begin{figure}[ht]
  \centering
\includegraphics[width=2.5in]{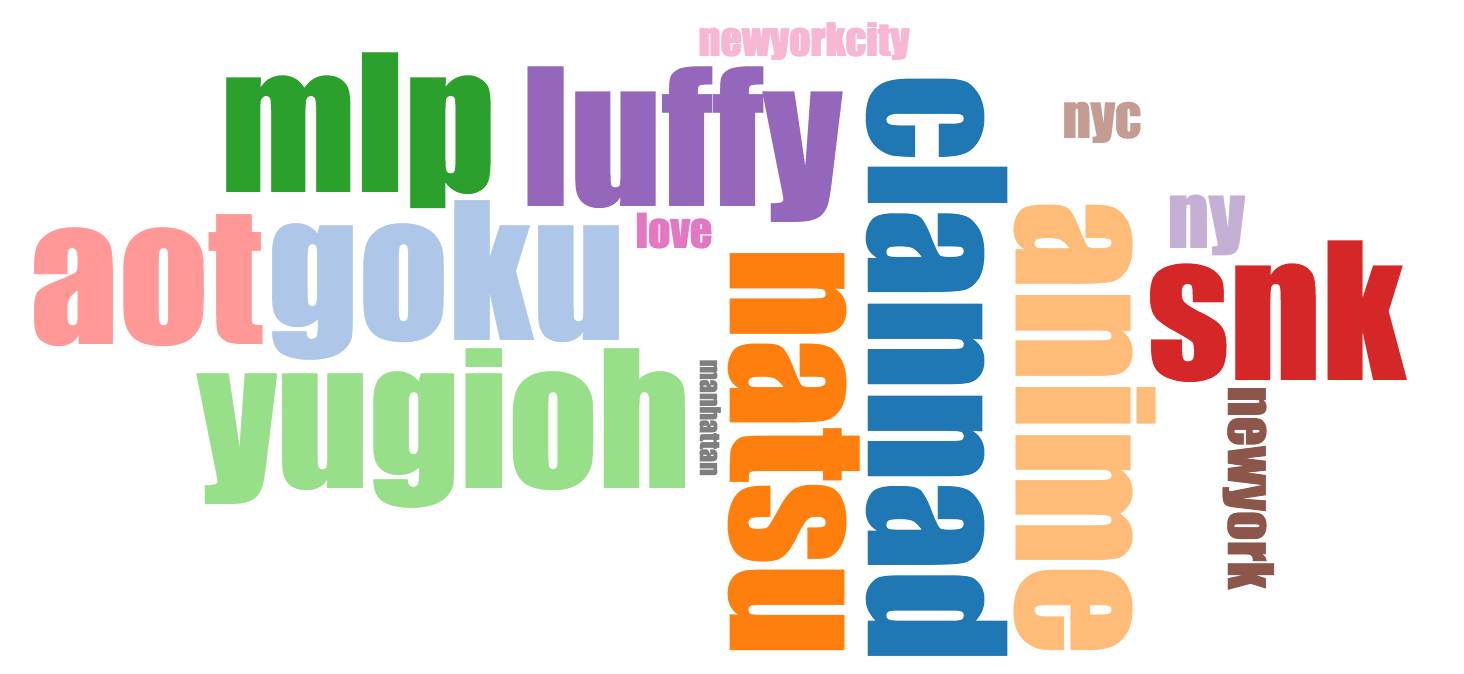}
  \caption{Illustration for NY}
  \label{like-filter}
\end{figure}

\section{Discussion}

\subsection{Results Discussion}

We present our work in the poster session. We are happy to see that all of our audiences find great pleasure in our work. Out of our expectation, they seem to be very interested in the word cloud. It is amazing to see what is happening in different cities, and it is quite interesting to see the different culture directly. They are also interested in the recommendation for advanced filters. Some people say they may apply the recommendation to become fancy.

However, they also express some confusion on the  design. Most people asked about the meaning of different colors on the map. This is reasonable, since the color is the most notable element in the design. They also asked about the meaning of the pie chart of the filters. We think this is because we do not provide enough text explanations. We have modified our final design  to meet the requirement of the audience.

\subsection{Future Work}
Since not all the photos are labeled with hashtags and not all the hashtags are correctly showing the content in each photo, using computer vision to analysis the real photo content, the style of the scenes and the major color theme may have stronger correlation with the filter types.

As the time changes people's visual preference may also changes, so the preference of filters may shifts as the time changes, we can learn the relationship with filters, likes and time to learn how visual preference changes and give out more current filter suggestion.

Since all the location analysis are based on the United States, so culture variety may be less between each cities, to extend the data to world based to learn some culture different between continents may give us more meaningful data. But world-wise spread of the Instagram usage may be the limit of this extension. In the future, we may also try to include weather, UV(Ultraviolet) \cite{zhang2013see}, and/or population information in addition to the location data. 

\section{Conclusion}

We have completed a visualization project on Instagram data. We study the relationship between the likes and the hashtags, location and filter, which many people are caring about. We use Tableau to visualize and analyze the raw data which has been processed to a easy-to use form and find interesting patterns. Then we use d3 to create a webpage interactive data visualization work. We combine multiple tools including datamap, aster plot and word cloud.

Compared with current existing Instagram analyzing tools, instead of focusing on individual user, we focus on big data on the whole Instagram community. We divide the data according to location, and in this way we detect culture different in different cities. This method can also be used in social science to study popular trend by collecting social network information. Also, we design a recommendation system, which is able to give general suggestions on choice of filters based on location. This may be improved with other advanced techniques like machine learning and computer vision techniques.

To summarize, it is amazing to visualize social network information data on real geographical map. This kind of work may be important in future development of data visualization and social science.

\balance{}

% REFERENCES FORMAT
% References must be the same font size as other body text.
\bibliographystyle{SIGCHI-Reference-Format}
\bibliography{proceedings}

\end{document}